\documentclass{article}
\pdfoutput=1
\usepackage[utf8]{inputenc}
\usepackage{url}
\usepackage{amsmath}
\usepackage{amssymb}
\usepackage{mathtools}
\usepackage{graphicx}
\usepackage{grffile} 
\usepackage[shortlabels]{enumitem}
\usepackage{algorithm}
\usepackage{setspace}
\usepackage{algorithmic,caption}
\usepackage{makecell, booktabs, array}

\usepackage{afterpage}

\usepackage{mathtools}
\usepackage{chngpage, comment}

\usepackage{tikz}
\usepackage[utf8]{inputenc}
\usepackage[english]{babel}
\usetikzlibrary{positioning, calc, shapes.geometric, shapes, shapes.multipart, arrows.meta, arrows, decorations.markings, external, trees}\usepackage[round]{natbib}
\usepackage[margin=1in]{geometry}
\usepackage{amsmath}
\usepackage{amssymb}
\usepackage{amsthm}

\usepackage{mathtools}

\newtheorem{theorem}{Theorem}

\title{\vspace{-1em}Bayesian Set of Best Dynamic Treatment Regimes and Sample Size Determination for SMARTs with Binary Outcomes }
\author{William J Artman, Ashkan Ertefaie, Kevin G Lynch, James R McKay}

\author{William J Artman\\[0.2cm]Department of Biostatistics and Computational Biology,\\ University of Rochester Medical Center,\\
Rochester, Saunders Research Building, 265 Crittenden Blvd., NY 14642, USA\\[0.1cm]
\url{William_Artman@URMC.Rochester.edu}\\[0.5cm] Ashkan Ertefaie\\[0.2cm]Department of Biostatistics and Computational Biology,\\ University of Rochester Medical Center,\\
Rochester, Saunders Research Building, 265 Crittenden Blvd., NY 14642, USA\\[0.5cm] Kevin G Lynch\\[0.2cm]
Center for Clinical Epidemiology and Biostatistics (CCEB)\\ and Department of Psychiatry,\\ University of Pennsylvania,\\ 3535 Market Street, 5099, Philadelphia, PA 19104, USA \\[0.5cm] James R McKay\\[0.2cm]Department of Psychiatry, Perelman School of Medicine,\\University of Pennsylvania,\\ 3535 Market St.,
Suite 500, Philadelphia, PA 19104, USA}
\date{}

\begin{document}
\maketitle

\def\ae{\color{blue}}

\newcommand{\btheta}{\boldsymbol\theta}
\newcommand{\bTheta}{\boldsymbol\Theta}
\newcommand{\bSigma}{\boldsymbol\Sigma}
\newcommand{\bsigma}{\boldsymbol\sigma}
\newcommand{\expit}{\mathrm{expit}}
\newcommand{\bD}{\boldsymbol D}
\newcommand{\bW}{\boldsymbol W}
\newcommand{\bZ}{\boldsymbol Z}
\newcommand{\by}{\boldsymbol y}
\newcommand{\bX}{\boldsymbol X}
\newcommand{\bu}{\boldsymbol u}
\newcommand{\bU}{\boldsymbol U}
\newcommand{\bbeta}{\boldsymbol \beta}
\newcommand{\var}{\mathrm{Var}}
\newcommand{\indep}{\rotatebox[origin=c]{90}{$\models$}}
\newcommand{\bigCI}{\mathrel{\text{\scalebox{1.07}{$\perp\mkern-10mu\perp$}}}}
\newcommand{\E}{\mathbb{E}}
\newcommand{\EDTR}{\mathrm{EDTR}}
\newcommand{\R}{\mathrm{R}}
\newcommand{\NR}{\mathrm{NR}}
\newcommand{\Odds}{\mathrm{Odds}}
\newcommand{\noopsort}[2]{#2}
\newcommand{\Cov}{\mathrm{Cov}}

\doublespacing
\begin{abstract}

One of the main goals of sequential, multiple assignment, randomized trials (SMART) is to find the most efficacious design embedded dynamic treatment regimes. The analysis method known as multiple comparisons with the best (MCB) allows comparison between dynamic treatment regimes and identification of a set of optimal regimes in the frequentist setting for continuous outcomes, thereby, directly addressing the main goal of a SMART. In this paper, we develop a Bayesian generalization to MCB for SMARTs with binary outcomes. Furthermore, we show how to choose the sample size so that the inferior embedded DTRs are screened out with a specified power. We compare log-odds between different DTRs using their exact distribution without relying on asymptotic normality in either the analysis or the power calculation. We conduct extensive simulation studies under two SMART designs and illustrate our method's application to the Adaptive Treatment for Alcohol and Cocaine Dependence (ENGAGE) trial.

\end{abstract}

\section{Introduction}
Substance use disorders are a common yet debilitating group of conditions for which treatment response is highly variable (\citealp{mckay2009continuing,kranzler2012personalized,black2014mechanisms,witkiewitz2015recommendations}). While there exist interventions, physicians often must rely on clinical experience to choose subsequent therapies for individuals who have failed to respond to initial treatment. In order to personalize medicine, there is a need for longitudinal trial data as well as methodologies for comparing sequences of treatment decision rules.

Sequential, multiple assignment, randomized trial (SMART) designs are a type of longitudinal clinical trial specifically designed to determine the optimal sequence of decision rules called dynamic treatment regimes (DTRs) (\citealp{LAVORI2000605,murphy2005experimental,lei2012smart,rose2019sample,murphy2001marginal,murphy2003optimal,robins2004optimal,nahum2012experimental,chakraborty2013statistical,chakraborty2014dynamic,laber2014dynamic}). In particular, there are DTRs embedded in the SMART by design; hence, SMARTs provide evidence for the regime tailored to a subject's response.

Standard multiple comparison approaches used to control for type I errors result in a loss
of statistical power. This is critically important when working with smaller data sets such as clinical trials.
\cite{ertefaie2015} introduced a frequentist semiparametric model for estimation of the embedded DTR outcomes as well as continuous outcome multiple comparisons with the best (MCB) for comparing the embedded DTRs. MCB enables the construction of a set of optimal embedded DTRs which are statistically indistinguishable for the available data. MCB offers increased power while controlling the type I error rate so that the best embedded DTR is included with a specified probability (\citealp{hsu1981simultaneous,hsu1984,hsu1996multiple,artman2018power}). By requiring only $L-1$ comparisons where $L$ is the number of embedded DTRs compared with all $\displaystyle \binom{L}{2}$ pairwise comparisons, MCB yields greater power over other methods. However, applications of MCB have focused on continuous outcomes. 

Binary outcomes arise frequently as primary outcomes in psychiatry and addiction clinical trials. The exiting methods for estimation in SMARTs with binary outcomes do not permit complex comparisons between all of the embedded DTRs such as with MCB \citep{kidwell2018design, dziak2019data}. Moreover, the power analysis approaches  for a SMART almost entirely focus on continuous outcomes aiming to size a SMART for simple comparisons between two embedded DTRs or to power a SMART so that the best embedded DTR has the largest point estimate with a specified probability \citep{CrivelloEvaluation2007,CrivelloStatMethod2007}.  More recently, \cite{rose2019sample} proposed a method which relies on Q-learning for continuous outcomes to size a SMART. 
\cite{kidwell2018design} developed a sample size calculator for binary outcome SMARTs, but their method is restricted to comparisons between only two embedded DTRs. Furthermore, they assume asymptotic normality which may not hold due to the small sample size of embedded treatment sequences in a SMART.

\cite{yan2020sample} developed a frequentist sample size calculation method for pilot SMARTs. In particular, they size a SMART so that the mean outcome of a DTR is within ``a margin of error''. In other words, so that confidence intervals are less than a pre-specified length. Advantages of their method include applicability to continuous, count, and binary outcomes. A limitation is that normality is assumed for all outcome summary statistics including log-OR. However, this may be unreasonable due to the small sample size in each branch of the SMART. This could be problematic especially in a pilot SMART for which the sample size is small by design. An advantage of our method is it does not make such strong parametric assumptions and uses uninformative priors to avoid bias. An important similarity of their method to ours is taking into account uncertainty in sample size calculations rather than only viewing the problem as that of estimation. In addition, for our method, the best embedded DTR will be included with a given probability. In summary, our method makes comparisons between $>2$ embedded DTRs adjusting for multiplicity, rather than looking at them marginally. The authors of \cite{yan2020sample} emphasize if there is no hypothesis testing in the pilot SMART, then standard power analysis is not relevant. Therefore, their method fills an important gap for pilot SMARTs. However, in this article, we are interested in drawing inference on the optimal embedded DTRs and sizing SMARTs for this purpose.

\cite{ogbagaber2016design} presented a method for power analysis based off an omnibus test as well as all pairwise comparisons for continuous outcomes. \cite{artman2018power} proposed a power analyses for SMARTs using MCB approach. Specifically, the latter method computes the
number of individuals to enroll in a SMART in order to achieve a specified power to exclude
embedded DTRs inferior to the best by a given amount from the set of best. This approach, however, relies on the normality assumption of the outcome, and thus, cannot always be used with binary outcomes. 

In this article, we extend MCB to the Bayesian binary outcome setting to directly address current methods' limitations. We build upon the work in \cite{artman2018power} to develop a rigorous power analysis procedure. We simulate from the posterior of each parameter for each treatment sequence individually which are independent. This circumvents the need for specifying the correlation matrix of the random embedded DTR outcomes in the frequentist setting. Then, using Robin's G-computation formula for a dynamic regime \citep{robins1986new}, we represent each embedded DTR outcome as a weighted average of the independent treatment sequences. We use the transformed draws in the MCB procedure to facilitate Bayesian inference. We leverage the method in \cite{mandel2008simultaneous} to extend the MCB methodology to use Monte Carlo draws in the construction of the set of best embedded DTRs. In particular, it enables the construction of simultaneous one-sided upper credible intervals for log-odds ratios, comparing each embedded DTR to the best embedded DTR when the outcome is binary. Our proposed approach
obviates many of the challenges associated with the current methods of choice,
namely
\begin{enumerate}[label=(\roman*),leftmargin=1cm]
  \item in contrast with the method of \cite{artman2018power}, our Bayesian MCB method can be used directly with binary outcomes and our power analysis requires fewer parameter specification;
  \item in contrast with the existing multiple comparison approaches, our approach performs fewer comparisons which increases the statistical power; and
  \item is designed specifically for statistical inference for complex comparisons of $>2$ embedded DTRs and is applicable to small sample size.
\end{enumerate}

In Section 2, we introduce the ENGAGE SMART. In Section 3, we formalize notation and present our Bayesian binary outcome model. In Section 4, we provide details about about Bayesian MCB. We then show how to perform power analysis. In Section 5, simulation studies are conducted. In Section  6, we illustrate our method on the the real ENGAGE SMART. In Section 7, we outline how to choose the input parameters for sample size determination. In Section 8, we conclude with discussion. The Appendix provides proofs.

\begin{figure}[t]
\centering
\includegraphics[width = 20cm,trim =5cm 5cm 0cm 0cm,clip=true]{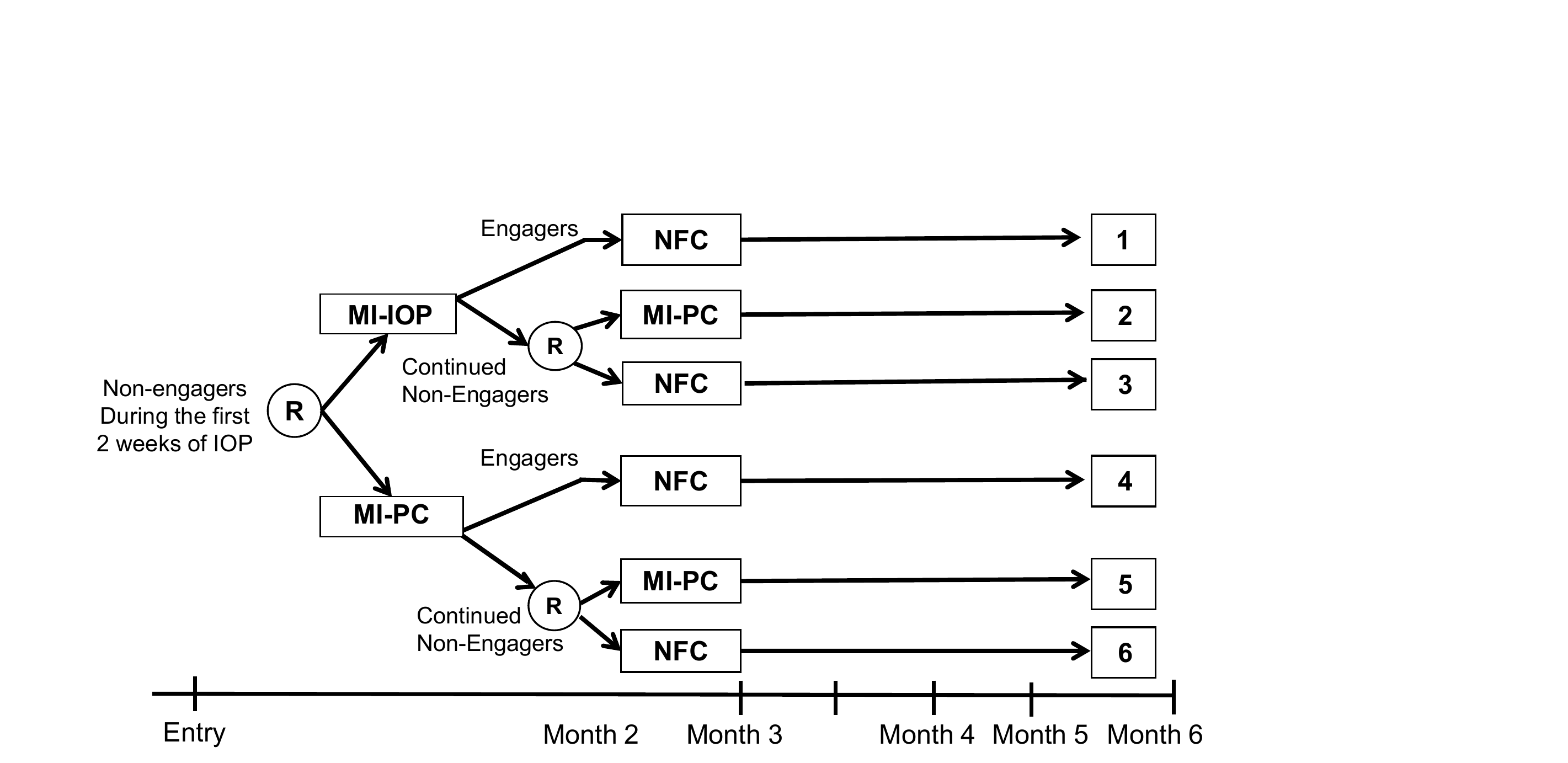}
\caption{Structure of the ENGAGE trial.}
\label{fig:ENGAGE-figure}
\end{figure}

 \begin{figure}[t]
\centering
\includegraphics[width = 7in, trim = 0cm 2cm 0cm 2cm,clip=true]{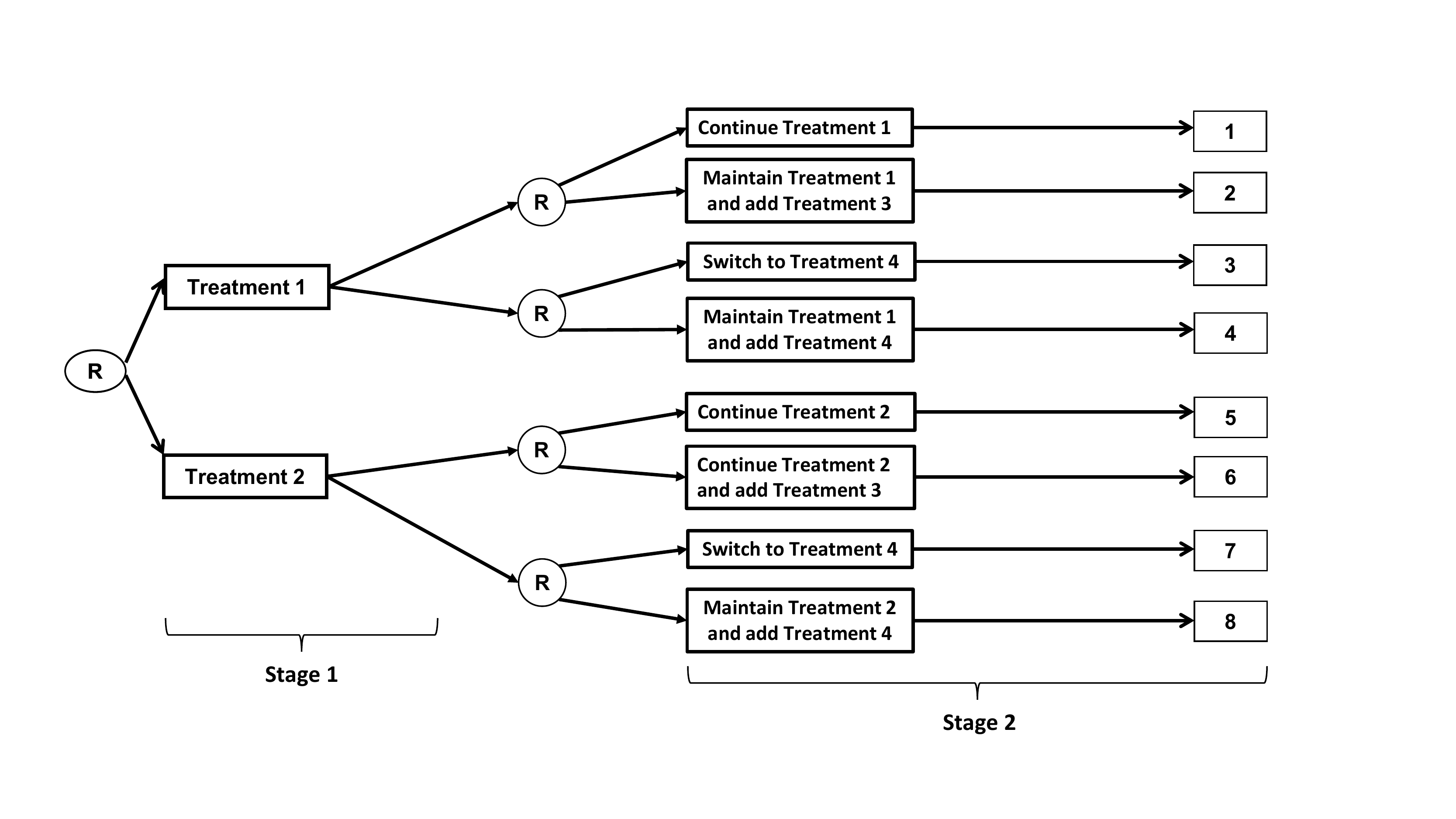}
\caption{Diagram of the General SMART.}
\label{General-SMART}
\end{figure}
\section{The ENGAGE trial}\label{SMARTs}

In the Adaptive Treatment for Alcohol and Cocaine Dependence SMART (ENGAGE; Figure \ref{fig:ENGAGE-figure}), individuals were initially randomized to two different treatments: motivational interviewing (MI) for intensive outpatient programs (IOP) and for patient's choice (PC). PC consisted of an intervention which the patient chooses. Patients who were non-engagers (did not attend any therapy sessions) at 2 weeks were enrolled. After 8 weeks, those who failed to attend any sessions during weeks 7 and 8 were classified as non-responders/non-engagers and were re randomized to either MI-PC or No Further Care (NFC). Those who were still responders were assigned to NFC. The original study analysis suggested that MI-IOP was significantly more effective than MI-PC. In this paper, we analyze the SMART to compare embedded DTRs. We restrict our attention in the study to individuals who did not engage by week 2 which yielded a sample size 148. One of the main goals of SMART designs is the identification of the best embedded DTRs for a primary outcome (\citealp{nahum2012experimental}). We construct a binary outcome which is whether the log of the total number of drinking days plus cocaine use days was less than the $25^{th}$ percentile of the log of the total number of drinking days plus cocaine use days.

Determination of optimal embedded DTRs would provide evidence-based recommendations for clinicians when treating individual patients with substance use disorders. The ENGAGE study aimed to determine the optimal treatment for non-responders (\citealp{mckay2015effect,van2015treatment}).
There were four embedded DTRs. Two of the embedded DTRs were as follows:
\begin{enumerate} 
\item Start with MI+IOP. If the subject is engaged 
during the first 8 weeks, then at the 8-week point, offer NFC; if the subject is labeled as a non-engager at week 8 of follow-up, offer MI+PC. 
\item Start with MI+IOP. If the subject is engaged during the first 8 weeks, then at the 8-week point, offer NFC; if the subject is labeled as a non-engager at week 8 of follow-up, offer NFC. 
\end{enumerate}
The other two embedded DTRs are similar (see Table  \ref{tab:EDTR-ENGAGE-table}). See \cite{mckay2015effect,van2015treatment} for more details about the SMART.

\section{Framework}

\subsection{Notation}
We focus on a SMART in which there are two stages where at each stage, individuals are randomized to different treatment options. By design, stage-2 treatment is commonly tailored based on a patient's ongoing response status which generates the different embedded DTRs (see Figures \ref{fig:ENGAGE-figure} and \ref{General-SMART}).
Let $Y^{(l)}$ denote the binary potential outcome for an individual following embedded DTR $l\in\{1,...,L\}$. Let $Y_k$ denote the binary potential outcome for an individual following treatment sequence $k \in \{1,...,K\}$. Let $A_1 \in \{-1,+1\}$ denote the stage-1 treatment assignment indicator. Let $A_2^{\mathrm{R}}\in\{-1,+1\}$ denote the stage-2 treatment assignment indicator for responders and $A_2^{\mathrm{NR}}\in\{-1,+1\}$ be the treatment assignment indicator for non-responders. Let $S_i$ denote the observed stage-1 treatment response indicator.
Let $\theta^{(l)}$ be the probability of response in the $l^{th}$ embedded DTR, $l=1,...,L$, where we assume that the L$^{th}$ embedded DTR is the best. Let $\theta_k$ denote the probability of response in the $k^{th}$ treatment sequence, $k=1,...,K$. Let $Z_i\in \{1,...,K\}$ denote the treatment sequence indicator for subject $i$.

 Then, the log-odds ratio for the $l^{th}$ embedded DTR compared with the best embedded DTR is $\zeta^{(l)} =\log\left(\dfrac{\Odds^{(l)}}{\Odds^{(L)}}\right)=\log(\Odds^{(l)})-\log(\Odds^{(L)})$ and $\Odds^{(l)}=\dfrac{\theta^{(l)}}{1-\theta^{(l)}}$. We do not make any distributional assumptions about the log-odds or the log-odds ratio. Assuming that being equal to 1 is the desirable outcome, higher log-odds (and equivalently, log-odds ratios) implies superiority of the corresponding embedded DTR.  
 \begin{table}[t]
\centering
\begin{tabular}[t]{cc}
\toprule
 & Embedded Dynamic Treatment Regime\\
\toprule
1 & $A_1=+1,S=1$ or $A_1=+1,S=0,A_2=+1$\\
&Receive MI-IOP, if respond receive NFC, if not respond switch to MI-PC\\
\midrule
2 & $A_1 = +1, S=1$ or $A_1=+1,S=0,A_2=-1$\\
&Receive MI-IOP, if respond, receive NFC, if not respond, receive NFC\\
\midrule
3 & $A_1=-1,S=1$ or $A_1=-1,S=0,A_2=+1$\\
&Receive MI-PC, if respond receive NFC, if not respond switch to MI-PC\\
\midrule
4 & $A_1=-1,S=1$ or $A_1=-1,S=0,A_2=-1$\\
&Receive MI-PC, if respond receive NFC, if not respond, receive NFC.\\
\bottomrule
\end{tabular}
\caption{Embedded dynamic treatment regime decision rules for the ENGAGE trial. MI-IOP: motivational interviewing for intensive outpatient program. MI-PC: motivational interviewing for patient choice. NFC: no further care.}
\label{tab:EDTR-ENGAGE-table}
\end{table}
\subsection{Binary outcome models}
  \allowdisplaybreaks
  We assume observations $Y_i^{\mathrm{obs}} \overset{iid}{\sim} \operatorname{Bern}(\theta_{Z_i})$ where $Z_i$ is the treatment sequence followed by the $i^{th}$ individual and $S_i \overset{iid}{\sim} \operatorname{Bern}(\lambda_{A_1})$ for $i =1,...,n$ where $A_1$ is the stage-1 treatment. 
  Our target parameter is the probability of response at the end of the trial ($Y^{(l)}=1$) for a particular embedded DTR. 
Using Robins' G-computation formula (\citealp{robins1986new}), we have the following representation for the target parameter: \begin{align}
 \Pr(Y^{(l)}=1)=
\Pr(Y=1\mid &A_{1}=a_{1l}, S=1,A_2^{\mathrm{R}}=a_{2l})\Pr(S=1\mid A_1=a_{1l})+\label{eqn:representation}\\
&\Pr(Y=1\mid A_1=a_{1l},S=0,A_2^{\mathrm{NR}}=a_{2l})\Pr(S=0\mid A_1=a_{1l}), \hspace{.1in} l=1,2,\cdots,L.\nonumber
\end{align}
Thus, the mean outcome under each embedded DTR is a weighted average of the responder treatment sequence outcome and the non-responder treatment sequence outcome. This permits estimation and inference without using a marginal structural model. While baseline covariates could be included in (\ref{eqn:representation}), for the purpose of power analysis, we define the target parameter as a marginalized quantity. In our Bayesian procedure, we impose a non-informative prior on $\theta_k$ and $\lambda_{A_1}$ which results in beta posterior distributions. We then define the treatment sequence response probabilities $\theta_{\mathrm{R}}=\Pr(Y=1\mid A_{1}=a_{1l}, S=1,A_2^{\mathrm{R}}=a_{2l}), \theta_{\mathrm{NR}}=\Pr(Y=1\mid A_1=a_{1l},S=0,A_2^{\mathrm{NR}}=a_{2l})$. Lastly, the probability of response in stage 1 for those assigned to $A_1$ is given by $\lambda_{A_1}=\Pr(S=1\mid A_1=a_{1}).$ \begin{table}[t]
\centering
\begin{tabular}[t]{cc}
\toprule
 & Embedded Dynamic Treatment Regime\\
\toprule
1 & $A_1=+1,S=1,A_2^{\mathrm{R}}=+1$ or $A_1=+1,S=0,A_2^{\mathrm{NR}}=+1$\\
&Receive Trt. 1, if respond, continue Trt 1, if not respond, switch to Trt 4.\\
\midrule
2 & $A_1 = +1, S=1,A_2^{\mathrm{R}}=+1$ or $A_1=+1,S=0,A_2^{\mathrm{NR}}=-1$\\
&Receive Trt. 1, if respond continue Trt 1, if not respond, add Trt. 4\\
\midrule
3 & $A_1=+1,S=1,A^{\mathrm{R}}_2=-1$ or $A_1=+1,S=0,A_2^{\mathrm{NR}}=+1$\\
&Receive Trt 1, if respond add Trt 3, if not respond switch to Trt 4\\
\midrule
4 & $A_1=+1,S=1,A^{\mathrm{R}}_2=-1$ or $A_1=+1,S=0,A_2^{\mathrm{NR}}=-1$\\
& Receive Trt. 1, if respond add Trt 3., if not respond add Trt. 4.\\
\midrule
5 & $A_1=-1,S=1,A_2^{\mathrm{R}}=+1$ or $A_1=-1,S=0,A_2^{\mathrm{NR}}=+1$\\
&Receive Trt. 2, if respond continue Trt. 2., if no respond switch to Trt 4.\\
\midrule
6 & $A_1 = -1, S=1,A_2^{\mathrm{R}}=+1$ or $A_1=-1,S=0,A_2^{\mathrm{NR}}=-1$\\
&Receive Trt. 2, if respond, continue Trt. 2, if not respond, add Trt 4.\\
\midrule
7 & $A_1=-1,S=1,A^{\mathrm{R}}_2=-1$ or $A_1=-1,S=0,A_2^{\mathrm{NR}}=+1$\\
& Receive Trt.2, if respond add Trt. 3, if not respond switch to Trt 4.\\
\midrule
8 & $A_1=-1,S=1,A^{\mathrm{R}}_2=-1$ or $A_1=-1,S=0,A_2^{\mathrm{NR}}=-1$\\
&Receive Trt. 2, if respond add Trt. 3., if not respond, add Trt 4.\\
\bottomrule
\end{tabular}
\caption{Embedded dynamic treatment regime decision rules for the General SMART.}
\label{tab:generalsmartEDTR}
\end{table} We leverage  (\ref{eqn:representation}) to transform each MCMC draw to the target parameter and, subsequently, into a log-odds ratio. By taking this simulation based approach, we avoid making normality assumptions. At the $m^{th}$ iteration of the MCMC, we draw $\theta_{\mathrm{R},m}$, $\theta_{\mathrm{NR},m}$, and  $\lambda_{A_1,m}$.
Hence, the $m^{th}$ draw of the probability of response for the $l^{th}$ embedded DTR is $\theta^{(l)}_m=\theta_{\mathrm{R},m}\lambda_{A_1,m}+\theta_{\mathrm{NR},m}(1-\lambda_{A_1,m}).$

The $m^{th}$ draw of the log-odds ratio is $\zeta^{(l)}_m=\log \mathrm{OR}^{(l)}=\log\left(\frac{\theta^{(l)}_m}{1-\theta^{(l)}_m}\right)-\log\left(\frac{\theta^{(L)}_m}{1-\theta^{(L)}_m}\right)$. Next, we specify the likelihoods. The likelihood for the observed end-of-study outcomes is given by the following $
    f(Y_1,...,Y_n\mid  \theta_1,...,\theta_k)
    \propto \prod_{i=1}^n \theta_{Z_i}^{Y_i}(1-\theta_{Z_i})^{1-Y_i},
$ where the treatment sequence indicator is $Z_i \in\{1,...,K\}$. We then assume a non-informative uniform prior on $\theta_{k}$, $k \in \{1,...,K\}$ which yields the posterior of $\theta_{k}$ given by the following: \begin{align*}
    f(\theta_{k}\mid y_1,...,y_n) &\sim \operatorname{Beta}\left(\sum_{i:Z_i=k}Y_i+1,\sum_{i:Z_i=k} 1-\sum_{i:Z_i=k}Y_i+1\right)
\end{align*}

Furthermore, the posterior of the stage-1 response probabilities are given by 
\begin{align*}
    f(\lambda_{A_1}\mid S_1,...,S_n)
    &= \operatorname{Beta}\left(\sum_{i:A_{1i}=A_1}S_i+1,\sum_{i:A_{1i}=A_1}1-\sum_{i:A_{1i}=A_1}S_i+1\right)
\end{align*}

\section{MCB for binary outcomes}\label{MCBBinary}
\subsection{Simultaneous credible intervals for Monte Carlo draws}
In this section, we describe our MCB procedure based off \cite{mandel2008simultaneous} to construct simultaneous $100(1-\alpha)\%$ one-sided upper credible intervals using multivariate draws from an arbitrary distribution. Their method was originally designed to construct confidence intervals for draws obtained using the bootstrap. In Section A.1. of the Appendix, we show that the latter approach is also applicable to Monte Carlo simulation draws of the posterior. The following theorem forms the basis for the extension of MCB to the Bayesian setting.

\begin{theorem}
Let $\zeta^{(l)}_m,l=1,\cdots,L$, $m=1,\cdots,M$ denote draws of the parameters $\zeta^{(l)}$ obtained from Monte Carlo simulation. Let $r(m,l)$ denote the rank of $\zeta^{(l)}_m$ and assume no ties. Then, $U^{(l)}=\zeta^{(l)}_{(r_{1-\alpha}l)},l=1,\cdots,L-1$ are simultaneous $1-\alpha$ level upper credible intervals for $\zeta^{(l)}, l=1\cdots,L-1$ where $r_{1-\alpha}$ is the $(1-\alpha)^{th}$ quantile of $\max_{l}r(m,l)$ for each draw $m=1,\cdots,M$.
\end{theorem}

The goal is to construct upper credible interval limits $U^{(l)}$, $l=1,\cdots,L$ which jointly have coverage $1-\alpha$. Note that there is a one-to-one correspondence between a draw $\zeta^{(l)}_m$ and its rank $r(m,l)$ in the absence of ties where $m$ is the $m^{th}$ draw, $m=1,\cdots,M$. In particular, $\zeta^{(l)}_{(r(m,l))}=\zeta^{(l)}_m$. Then, $\zeta^{(l)}_m$ is less than $U^{(l)}$ for all $l=1,\cdots,L$ for $100(1-\alpha)\%$ of the draws $m=1,\cdots,M$ if and only if $\zeta^{(l)}_{(r(m,l))}$ is less than $U^{(l)}$ for all $l=1,\cdots,L$ for $100(1-\alpha)\%$ of the draws $m=1,\cdots,M$.
A sufficient condition is for $r(m,l)$ to be less than some integer  $r_{1-\alpha}$ for all $l$, for $100(1-\alpha)\%$ of the $r(m,l)$, $l=1,\cdots,L,\ m= 1,\cdots,M$. Equivalently, the max rank across embedded DTRs $r(m):=\max_l r(m,l)$ is less than $r_{1-\alpha}$ for $100(1-\alpha)\%$ of the draws $m=1,\cdots,M$. Let $r_{1-\alpha}$ equal the $1-\alpha$ quantile of $r(1),\cdots,r(M)$. Then, letting $U^{(l)} = \zeta^{(l)}_{(r_{1-\alpha}l)}$ for $l=1,\cdots,L$ completes the proof. See the Appendix for more details. The proof that the algorithm works assumes no ties. When there are ties in rank, one can use the minimum rank to obtain coverage near $100(1-\alpha)\%$ for a sufficiently large number of draws (\citealp{mandel2008simultaneous}).

Define $\hat{\mathcal{B}}=\{\EDTR^{(l)} \mid U^{(l)}\geq 0 \}$ where $U^{(l)}$ is the upper credible limit for the difference between the $l^{th}$ embedded DTR's log-odds and the best embedded DTR's log-odds. Then, regardless of sample size, $\hat{\mathcal{B}}$ contains the true optimal embedded DTR with probability at least $1-\alpha$ by construction. It furthermore adjusts for multiplicity by the above procedure. We define $\hat{\mathcal{B}}$ to be the set of best embedded DTRs as it contains embedded DTRs which are not statistically significantly inferior to the best embedded DTR. This is analogous to MCB constructed for continuous outcomes in \cite{ertefaie2015}. Sample sizes which are larger shrink $\hat{\mathcal{B}}$ so that inferior embedded DTRs are excluded. In the next section, we prove how to choose the sample size to achieve this goal.

\subsection{Power calculation with binary outcomes}\label{powercompbinary}
In the last section, we demonstrated how to construct a set of optimal embedded DTRs. In this section, we show how to size a SMART to obtain a set of embedded DTRs which are not different in their efficacy by a clinically meaningful amount. In terms of the credible intervals, those which cover 0 correspond with embedded DTRs which are not statistically distinguishable from the optimal embedded DTR. Those that are strictly below 0 are inferior and are subsequently excluded from the set of best. We present how to perform power analysis for Bayesian MCB with a binary outcomes.

Let $\boldsymbol{\eta}$ be the known inputs (response probabilities at each stage). We wish to determine the sample size $n$ such that $$\Pr\left(\bigcap_{l:\Delta_l\geq\Delta} \left\{U^{(l)}<0\right\}\mid \boldsymbol{\eta}\right)=1-\gamma$$ where $1-\gamma$ is the power, $\Delta$ is the clinically meaningful difference with the best embedded DTR and $\Delta_l$ is the log-odds ratio between the $l^{th}$ embedded DTR and the best embedded DTR, 
Without loss of generality, assume that $Y^{\mathrm{new}}$ is a sufficient statistic for $\boldsymbol{\eta}$. Then, 
\begin{align*}
    &\Pr\left(\bigcap_{l:\Delta_l\geq\Delta} \left\{U^{(l)}<0\right\}\mid \boldsymbol{\eta}\right)=\int \Pr\left(\bigcap_{l:\Delta_l\geq\Delta} \left\{U^{(l)}<0\right\}\mid Y^{\mathrm{new}},\boldsymbol{\eta}\right)p(Y^{\mathrm{new}}\mid\boldsymbol{\eta})dY^{\mathrm{new}}\\
    &=\int \int \Pr\left(\bigcap_{l:\Delta_l\geq\Delta} \left\{U^{(l)}<0\right\}\mid Y^{\mathrm{new}},\boldsymbol{\eta},\phi^{(1)},...,\phi^{(M)}\right)p(\phi^{(1)},...,\phi^{(M)}\mid Y^{\mathrm{new}})p(Y^{\mathrm{new}}\mid \boldsymbol{\eta})d\boldsymbol{\phi}dY^{\mathrm{new}},
\end{align*}
where $\phi^{(1)},...,\phi^{(M)}$ are independent of $\boldsymbol{\eta}$ given $Y^{\mathrm{new}}$ by sufficiency and are  draws of the parameters for which we are constructing upper credible intervals. 
Therefore,
\begin{align*}
    &\int \int \Pr\left(\bigcap_{l:\Delta_l\geq\Delta} \left\{U^{(l)}<0\right\}\mid Y^{\mathrm{new}},\boldsymbol{\eta},\phi^{(1)},...,\phi^{(M)}\right)p(\phi^{(1)},...,\phi^{(M)}\mid Y^{\mathrm{new}})p(Y^{\mathrm{new}}\mid \boldsymbol{\eta})d\boldsymbol{\phi}dY^{\mathrm{new}}\\
    &=\int \int \Pr\left(\bigcap_{l:\Delta_l\geq\Delta} \left\{U^{(l)}<0\right\}\mid \phi^{(1)},...,\phi^{(M)}\right)p(\phi^{(1)},...,\phi^{(M)}\mid Y^{\mathrm{new}})p(Y^{\mathrm{new}}\mid \boldsymbol{\eta})d\boldsymbol{\phi}dY^{\mathrm{new}}\\
    &=\int \int \mathcal{I}\left(\bigcap_{l:\Delta_l\geq\Delta} \left\{U^{(l)}(\phi^{(1)},...,\phi^{(M)})<0\right\}\right)p(\phi^{(1)},...,\phi^{(M)}\mid Y^{\mathrm{new}})p(Y^{\mathrm{new}}\mid \boldsymbol{\eta})d\boldsymbol{\phi}dY^{\mathrm{new}}\\
    &\approx\dfrac{1}{I_1}\sum_{i=1}^{I_1} \int \mathcal{I}\left(\bigcap_{l:\Delta_l\geq\Delta} \left\{U^{(l)}(\phi^{(1),i},...,\phi^{(M),i})<0\right\}\right)p(\phi^{(1),i},...,\phi^{(M),i}\mid Y^{\mathrm{new},i})d\boldsymbol{\phi},
    \end{align*}
where $U^{(l)}(\cdot,...,\cdot)$ maps the posterior draws to the $k^{th}$ simultaneous credible interval as determined by the procedure in the previous section. Furthermore, $Y^{\mathrm{new},i}\sim p(Y^{\mathrm{new}}\mid \boldsymbol{\eta})$ for all $i=1,...,I_1$. In the above, we have used the Law of Large Numbers. Similarly, 
\begin{align*}
    &\dfrac{1}{I_1}\sum_{i=1}^{I_1} \int \mathcal{I}\left(\bigcap_{l:\Delta_l\geq\Delta} \left\{U^{(l)}(\phi^{(1),i},...,\phi^{(M),i}<0\right\}\right)p(\phi^{(1),i},...,\phi^{(M),i}\mid Y^{\mathrm{new},i})d\boldsymbol{\phi}\\
    &\approx\dfrac{1}{I_1}\sum_{i=1}^{I_1}\dfrac{1}{J_1}\sum_{j=1}^J \mathcal{I}\left(\bigcap_{l:\Delta_l\geq\Delta} \left\{U^{(l)}(\phi^{(1),i}_j,...,\phi^{(M),i}_j)<0\right\}\right),
\end{align*}
 where $\phi_j^{(m),i}\sim p(\phi\mid Y^{\mathrm{new},i})$ and $Y^{\mathrm{new},i}\sim p(Y^{\mathrm{new}}\mid\boldsymbol{\eta})$.
 In summary,    
\begin{align*}
     \Pr\left(\bigcap_{l:\Delta_l\geq\Delta} \left\{U^{(l)}<0\right\}\mid \boldsymbol{\eta}\right)
     &\approx\dfrac{1}{I_1}\sum_{i=1}^{I_1}\dfrac{1}{J_1}\sum_{j=1}^J \mathcal{I}\left(\bigcap_{l:\Delta_l\geq\Delta} \left\{U^{(l)}\left(\phi^{(1),i}_j,...,\phi^{(M),i}_j\right)<0\right\}\right).
\end{align*}
    
Therefore, we may choose a sample size that yields $100(1-\gamma)\%$ power by the following procedure.

Specify a grid of sample sizes $n \in \{ n_{min},...,n_{max}\}$. For each $n$ in the grid, compute the statistical power using steps 2-4 and for each $n$ in the grid calculate the power.

\begin{enumerate}
    \item Draw $Y^{\mathrm{new},i}\sim p(Y^{\mathrm{new}}\mid\boldsymbol{\eta}),i=1,...,I_1$ where the input parameters $\boldsymbol{\eta}$ are given by the stage-1 treatment response probabilities and  end-of-study response probabilities.
    \item Draw $\phi^{(m),i}_j\sim p(\phi\mid Y^{\mathrm{new},i})$ for all $i=1,...,I_1$ and $j=1,...,J_1$, $m=1,...,M$.
    \item Compute $U^{(l)}\left(\phi^{(1),i}_j,...,\phi^{(M),i}_j\right)$, $l=1,...,L$.
    \item Iterate steps 2-4, $M$ (e.g., $M=500$) times and repeat until $n$ is sufficiently large so that\\ $$\dfrac{1}{I_1}\sum_{i=1}^{I_1}\dfrac{1}{J_1}\sum_{j=1}^J \mathcal{I}\left(\bigcap_{l:\Delta_l\geq\Delta} \left\{U^{(l)}(\phi^{(1),i}_j,...,\phi^{(M),i}_j)<0\right\}\right)\geq 1- \gamma.$$
\end{enumerate}

\section{Simulation studies}\label{sim}
\subsection{SMART design 1 simulation study}

See Figure \ref{fig:ENGAGE-figure} for a flow chart of the simulation design 1 SMART. Design 1 includes 6 embedded treatment sequences and 4 embedded DTRs. People who respond to their stage-1 treatment continue on the same treatment, whereas non-responders are re-randomized to one of two rescue treatments. The following is the simulation scheme:
\begin{enumerate}
    \item Stage-1 treatment assignment: $A_1\in\{-1,1\}$ with $\mathrm{probability}=0.5$.

    \item Stage-1 treatment response indicator: $S\sim\mathrm{Bern}(\lambda_{A_1})$, $A_1 \in \{-1,+1\}$.
    \item Stage-2 treatment for non-responders: $A_2 \in\{-1,1\}$ with $\mathrm{probability}=0.5$.
    \item Final binary outcome: $Y_{Z_i} \sim \mathrm{Bern}(\pi_{Z_i}), i=1,...,n$.
\end{enumerate}

\subsection{General SMART simulation study}
See Figure \ref{General-SMART} for a flow chart of the simulation general SMART. The general SMART has 8 embedded treatment sequences and 8 embedded DTRs. The following is the simulation scheme:
\begin{enumerate}
    \item Stage-1 treatment assignment: $A_1 \in \{-1,1\}$ with $\mathrm{probability}=0.5$.
    \item Stage-1 treatment response indicator: $S\sim\mathrm{Bern}(\lambda_{A_1})$, $A_1 \in\{-1,+1\}$.
    \item Stage-2 treatment for responders to stage-1 treatment ($S=1$): $A_2^{\operatorname{R}}\in \{-1,1\}$ with $\mathrm{probability}=0.5$.
    \item Stage-2 treatment for non-responders to stage-1 treatment ($S=0$): $A_2^{\operatorname{NR}} \in \{-1,1\}$ with $\mathrm{probability}=0.5$.
    \item The final binary outcome is: $Y_i \sim \mathrm{Bern}(\pi_{Z_i}), i= 1,...,n$.
\end{enumerate}

One could incorporate covariates in either simulation scheme, but this would impose modeling assumptions which may lead to misspecification if, for example, probit or logistic regression is employed. Furthermore, including covariates would complicate the power analysis procedure as well as the interpretation of MCB. Hence, we directly simulate the variables based off the probabilities of response.

\begin{center}
\begin{table*}[t]%
\centering
\begin{tabular*}{400pt}{@{\extracolsep\fill}lcccccccccccc@{\fill}}
\toprule
&\multicolumn{4}{@{}c@{}}{\textbf{$n=100$}} & \multicolumn{4}{@{}c@{}}{\textbf{$n=400$ }} \\
\cmidrule{2-5}\cmidrule{6-9}\\
&\multicolumn{2}{@{}c@{}}{\textbf{Bayesian}} &\multicolumn{2}{@{}c@{}}{\textbf{MSM}}&\multicolumn{2}{@{}c@{}}{\textbf{Bayesian}}&\multicolumn{2}{@{}c@{}}{\textbf{MSM}}\\\cmidrule{2-3}\cmidrule{4-5}\cmidrule{6-7}\cmidrule{8-9}\\
\toprule
\textbf{Design 1} & \textbf{Bias} &\textbf{SD} & \textbf{Bias}&\textbf{SD}  & \textbf{Bias}  & \textbf{SD} & \textbf{Bias} & \textbf{SD}  \\
\midrule
$\theta_1$ & 0.0024 & 0.0725 & 0.0025 & 0.0819 & 0.0028 & 0.0397 & 0.0031 & 0.0409\\
$\theta_2$ & 0.0043 & 0.0722 & 0.0051 & 0.0816 & 0.0010 & 0.0393 & 0.0010 & 0.0404\\
$\theta_3$ & 0.0204 & 0.0730 & 0.0061 & 0.0822 & 0.0045 & 0.0397 & 0.0026 & 0.0409\\
$\theta_4$ & 0.0080 & 0.0768 & 0.0028 & 0.0874 & 0.0024 & 0.0441 & 0.0006 & 0.0461\\
\bottomrule
\textbf{General} & \textbf{Bias} &\textbf{SD} & \textbf{Bias}&\textbf{SD}  & \textbf{Bias}  & \textbf{SD} & \textbf{Bias} & \textbf{SD}  \\
\midrule
$\theta_1$ & 0.0172 & 0.0833 & 0.0025 & 0.1002 & 0.0027 & 0.0475 & 0.0054 & 0.0513\\
$\theta_2$ & 0.0038 & 0.0845 & 0.0025 & 0.1017 & 0.0021 & 0.0494 & 0.0038 & 0.0533\\
$\theta_3$ & 0.0432 & 0.0703 & 0.0059 & 0.0768 & 0.0114 & 0.0398 & 0.0041 & 0.0389\\
$\theta_4$ & 0.0297 & 0.0769 & 0.0072 & 0.0859 & 0.0107 & 0.0432 & 0.0022 & 0.0440\\
$\theta_5$ & 0.0194 & 0.0812 & 0.0011 & 0.0976 & 0.0067 & 0.0440 & 0.0019 & 0.0471\\
$\theta_6$ & 0.0239 & 0.0802 & 0.0001 & 0.0942 & 0.0072 & 0.0428 & 0.0005 & 0.0454\\
$\theta_7$ & 0.0183 & 0.0781 & 0.0004 & 0.0951 & 0.0052 & 0.0442 & 0.0007 & 0.0481\\
$\theta_8$ & 0.0228 & 0.0772 & 0.0015 & 0.0915 & 0.0057 & 0.0435 & 0.0007 & 0.0455\\
\bottomrule
\end{tabular*}
\caption{Simulation SMART designs with 4 embedded DTRs and one with 8, respectively. Embedded DTR performance estimates are given using our Bayesian approach compared with ``weighted and replicated'' logistic regression with a marginal structural model $m(\bbeta,\boldsymbol{A})=\beta_0+\beta_1A_1+\beta_2A_2^{\mathrm{R}}+\beta_3A_2^{\mathrm{NR}}+\beta_4A_1A_2^{\mathrm{R}}+\beta_5 A_1A_2^{\mathrm{NR}}$  (\citealp{kidwell2018design, nahum2012experimental, almirall2014introduction}).\label{estimation}}
\end{table*}
\end{center}
\begin{figure}[t]
\centering
\includegraphics[width = 7cm,trim =0cm 0cm 0cm 0cm,clip]{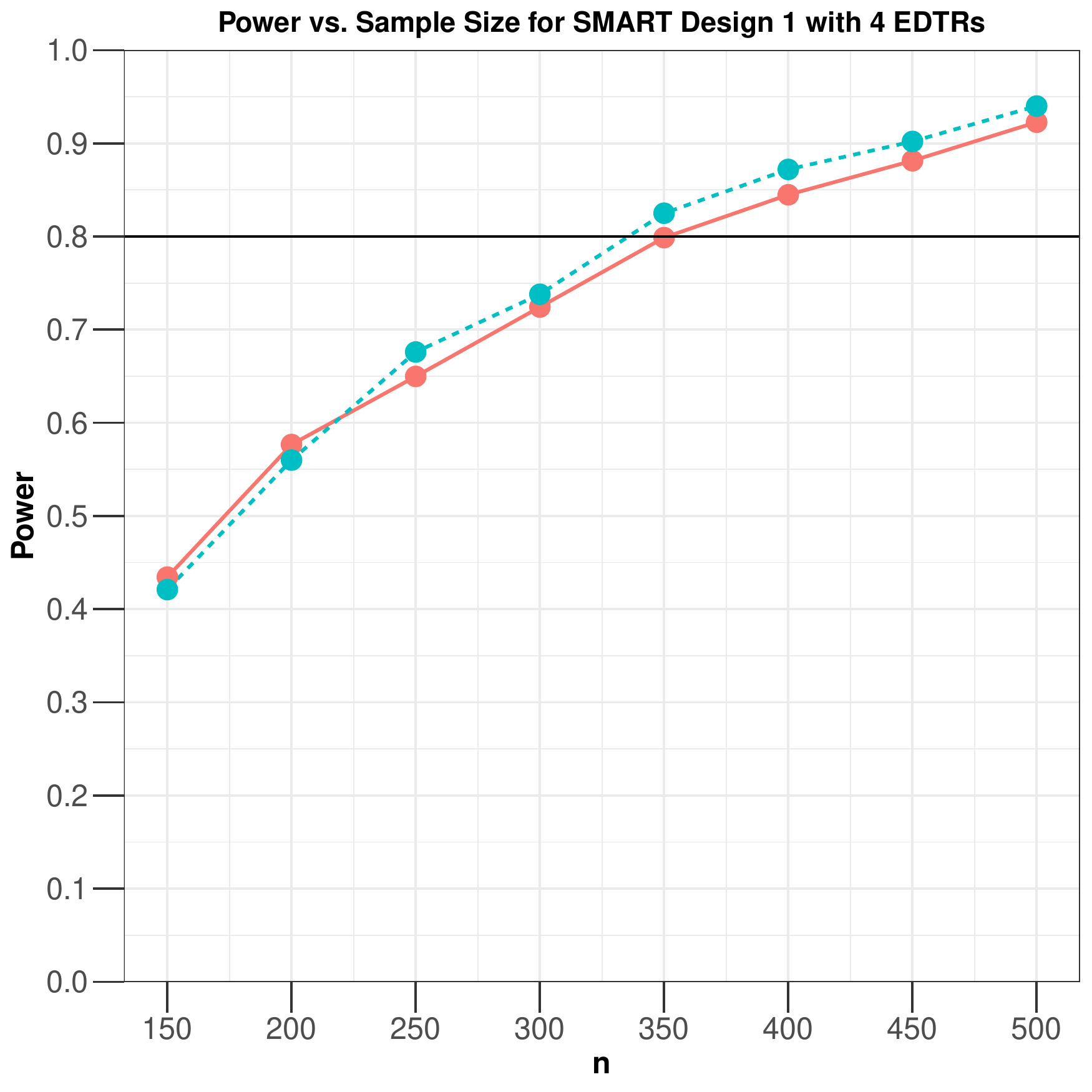}
\includegraphics[width = 7cm,trim =0cm 0cm 0cm 0cm,clip]{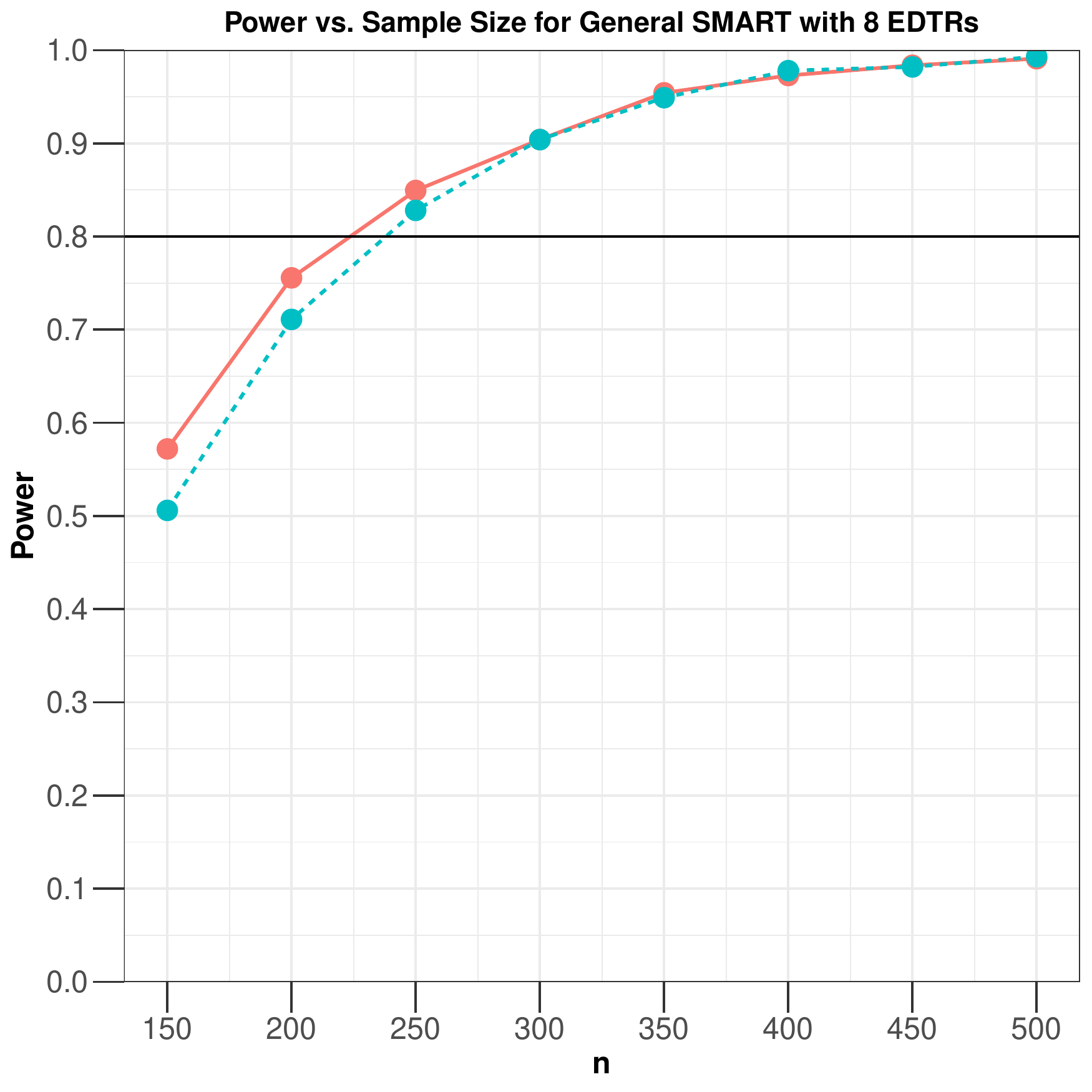}
\includegraphics[width = 7cm,trim =0cm 0cm 0cm 0cm,clip]{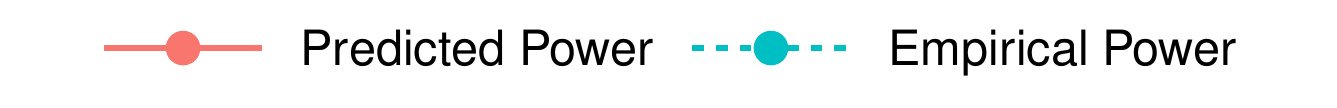}
\caption{Power vs. sample size plots for SMART simulations with 4 and 8 embedded DTRs, respectively. The predicted power is close to the empirical power. The black horizontal line corresponds to 80\% power.}
\label{fig:power-plots}
\end{figure}
\afterpage{
\begin{figure}[t]
\centering
\includegraphics[width = 12cm,trim =0cm 0cm 0cm 0cm,clip]{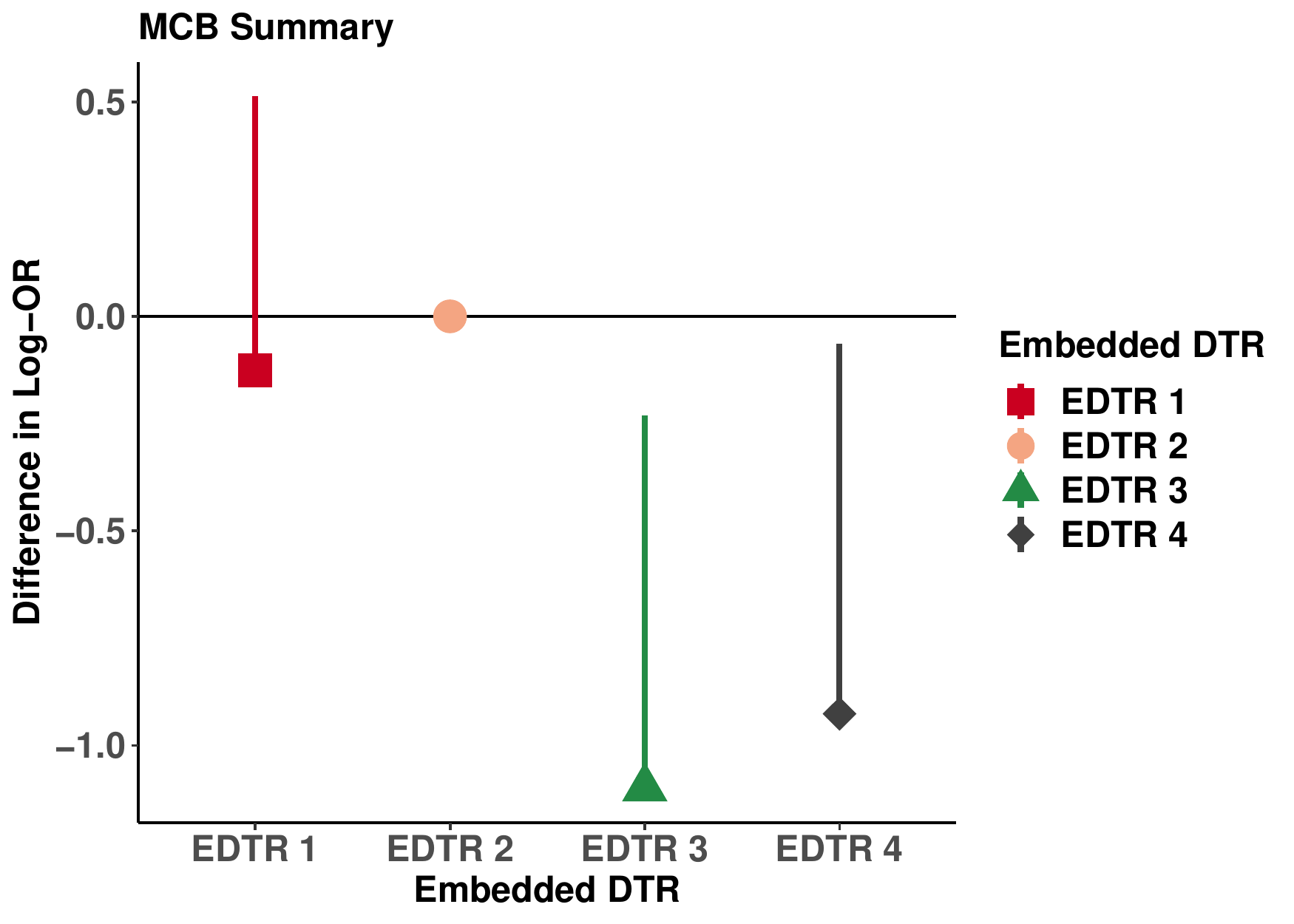}
\caption{Real ENGAGE SMART MCB. Upper one-sided credible intervals for the difference between each embedded DTR (EDTR) and the best embedded DTR. Intervals covering 0 indicate that the corresponding embedded DTR is not statistically significantly inferior to the best embedded DTR. Being below zero indicates inferiority compared with the best.}
\label{fig:MCB}
\end{figure}
}
\subsection{Simulation: results}
Table \ref{estimation} shows the bias and SD for our Bayesian estimation approach and for a frequentist approach weighted and replicated logistic regression with the marginal structural model (MSM)  $$m(\bbeta,\boldsymbol{A})=\beta_0+\beta_1A_1+\beta_2A_2^{\mathrm{R}}+\beta_3A_2^{\mathrm{NR}}+\beta_4A_1A_2^{\mathrm{R}}+\beta_5 A_1A_2^{\mathrm{NR}}.$$ We see that the bias is similar between the Bayesian procedure and MSM. The Bayesian procedure has slightly smaller SE.

To compute the empirical power, we simulated $1000$ datasets over a grid of sample sizes, $150,200,...,500$. We used $1000$ Monte Carlo draws for each of the $1000$ datasets to obtain $1000$ upper credible interval limits for each embedded DTR, e.g., $U_i^{(l)}, l=1,...,L$ and $i =1,...,1000$ for each embedded DTR $i$ and each sample size. We then computed the empirical power as $$\mathrm{Power}\approx\dfrac{1}{1000}\sum_{i=1}^{1000} \mathcal{I}\left(\bigcap_{l:\Delta_l\geq\Delta_{\mathrm{min}}}\left\{U^{(l)}_i<0\right\}\right)$$ where $\Delta_l = \log\left(\dfrac{\Odds_L}{\Odds_l}\right)$ where $\Delta_{\mathrm{min}}$ was set to $\Delta_{\mathrm{min}}=0.61$ for Design 1 with $\boldsymbol\Delta = (0.59, 1.30, 0.67, 0.00)^{\top}$. Then, $\EDTR_2$ and $\EDTR_3$ should be excluded from the set of best.

We chose $\Delta_{\mathrm{min}}=0.9$ with $\boldsymbol\Delta = (0.93, 1.93, 0.00, 1.14, 2.66, 1.94, 0.84, 0.17)^{\top}$. Then, $\EDTR_2,\EDTR_4,$ $\EDTR_5,\EDTR_6$, and $\EDTR_8$ should be excluded from the set of best. The power plots (Figure \ref{fig:power-plots}) demonstrate the accuracy of the predicted power estimating the required sample size.

\section{Constructing a set of best embedded DTRs for the ENGAGE SMART}\label{ENGAGE}

In this section, we apply our method to the The Adaptive Treatment for Alcohol and Cocaine Dependence (ENGAGE) SMART trial to construct a set of optimal embedded DTRs for a binary outcome. We dichotomize the log of the sum of the total number of days drinking plus the total number of days using cocaine plus a small constant 0.5 and consider a positive response to be having the outcome less than the 25th percentile. We see that the optimal embedded DTRs are embedded DTRs 1 and 2. The one with the lowest point estimate is embedded DTR 2. This suggests that MI-IOP is superior to patients' choosing their own care. Furthermore, although not statistically significant different, stage-2 NFC has a greater point estimate compared to MI-PC. The probability of response in each embedded DTR was $(\EDTR_1,\EDTR_2,\EDTR_3,\EDTR_4)=(0.38, 0.41, 0.19, 0.22)$.

The actual power is approximately 57\% for a sample size of 148, the actual sample size in the SMART. To  achieve a power of 80\% would require a sample size of approximately 250 subjects and for 90\%, approximately 350 subjects. 
The proposed MCB analysis demonstrates that the embedded DTR for which there is no further care are not statistically distinguishable from the embedded DTR for which patient choice is offered in stage 2. Therefore, clinicians may choose from the set of best either embedded DTR considering other factors such as costs, treatment burden and patients preference.
\begin{figure}[t]
\centering
\includegraphics[width = 8cm,trim =0cm 0cm 0cm 0cm,clip=true]{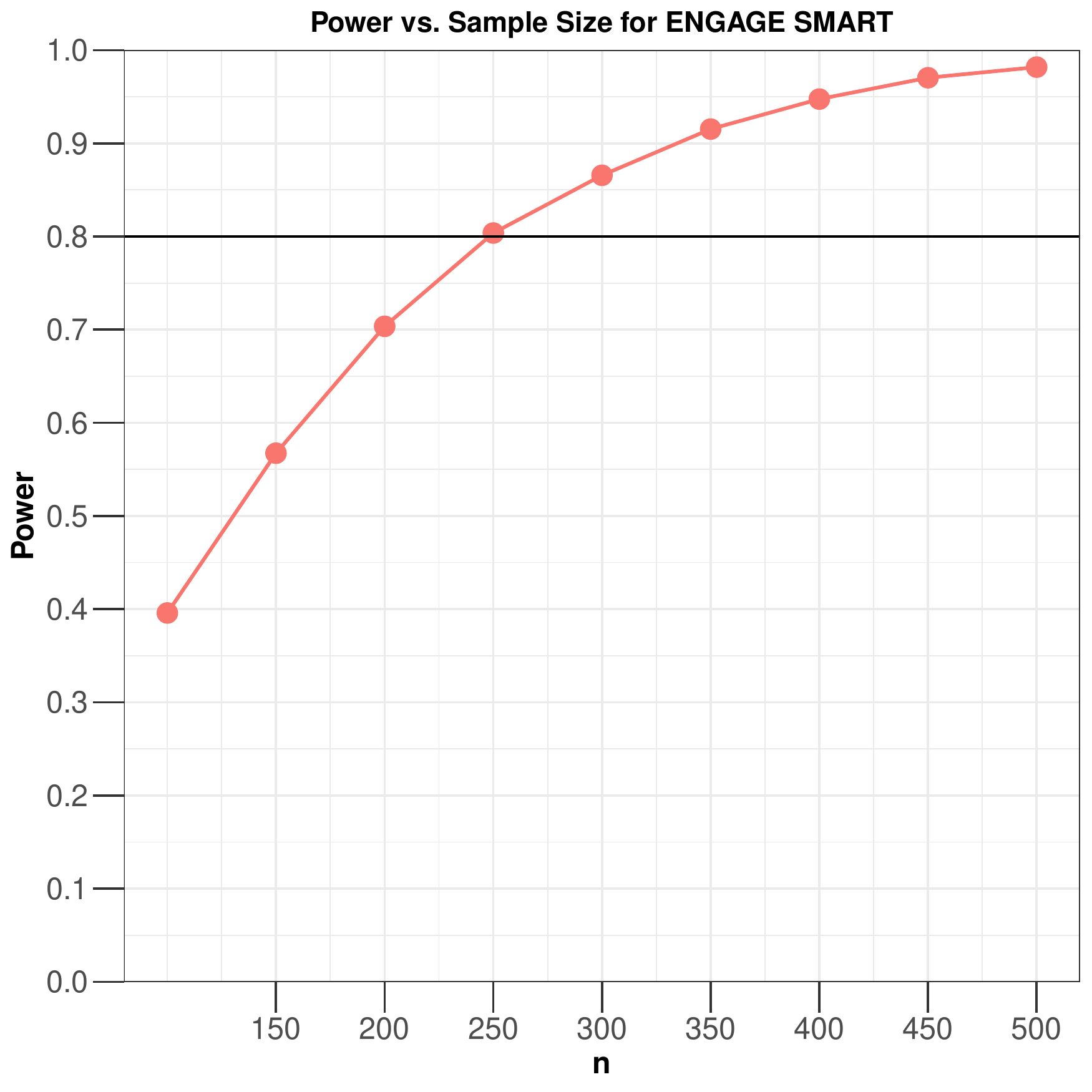}
\caption{Power for ENGAGE SMART to exclude all embedded DTRs except for the third (MI-PC stage 1, MI-PC stage 2). The actual power is 57\%.}
\label{fig:ENGAGE-power-plot}
\end{figure}
\section{Input parameters for sample size calculation}\label{inputparam}
The input parameters are the final response rates for each embedded treatment sequence $\theta_{k}$ for $k=1,...,K$, and the stage-1 treatment response probabilities $\lambda_{A_1}$ for $A_1 \in \{-1,+1\}$.
For the treatment sequences in which patients responded to the first stage treatment (probability $\lambda_{A_1}$) when continuing initial treatment, the response rate from non-SMART studies may be used. For sequences involving two sequential treatments, the response rates of the second stage treatment will yield a conservative estimate of the power, so in the absence of data on a particular sequence of treatments, one may choose the response rate of the second stage treatment obtained from existing literature.  Furthermore, one may equate different branch response probabilities essentially making the assumption of no interaction between stage-1 treatment and the response rate for a particular stage-2 treatment. This may be reasonable for a SMART in which both branches have the same treatment options in stage 2. As an alternative, one may conduct a pilot SMART to estimate the response rates \citep{artman2018power, rose2019sample}. 

\section{Discussion}\label{discussion}
 
An important goal of SMARTs is the determination of the best embedded DTRs. Subjects with substance use disorders are heterogeneous in their response to outcome. The optimal embedded DTRs for the ENGAGE study were those which consisted of the intensive out patient programs. Previous work has focused on inference for simple comparisons such as those between two embedded DTRs. Furthermore, methods which are applicable to more complex comparisons such as determination of the optimal embedded DTR while taking into account uncertainty has focused on continuous outcomes in the frequentist setting. We made two main contributions. First, we extended existing methodologies from continuous to binary outcomes and from the frequentist to the Bayesian setting. In particular, we applied the algorithm in \cite{mandel2008simultaneous} to MCB to construct a set of optimal embedded DTRs. Secondly, we provide a means for sample size calculations without the need for knowledge of the covariance matrix of log-odds ratios and without normality assumptions.

 We proved the validity of our method and demonstrated its application on real data, the ENGAGE SMART. Recently, there has been work to develop sample size formula based off Q-learning for a continuous outcome in \cite{rose2019sample}. It would be of interest to extend the proposed Bayesian MCB framework to Q-learning with binary outcomes.
  Furthermore, it would be valuable to extend MCB to constructing a set of best for longitudinal outcomes or zero-inflated outcomes. It would be straightforward to extend the Bayesian approach to encompass continuous outcomes.

\bibliographystyle{authordate1}
\bibliography{sample}
\clearpage

\appendix
\section{Proofs}
\subsection{Robin's G-Computation formula}\label{appendixA}
\begin{align*}
    \Pr(Y^{(l)}=1) &= \Pr(Y=1 \mid A_1,\EDTR^{(l)})\\
    &=\Pr(Y=1\mid A_1,S=1,\EDTR^{(l)})\Pr(S=1\mid A_1,\EDTR^{(l)})\\&+\Pr(Y=1\mid A_1,S=0,\EDTR^{(l)})\Pr(S=0\mid A_1,\EDTR^{(l)})\\
    &=\Pr(Y=1\mid A_{1}, S=1)\Pr(S=1\mid A_1)+\Pr(Y=1\mid A_1,S=0,A_{2}^{\mathrm{NR}})\Pr(S=0\mid A_1)
\end{align*}

\subsection{Proof of credible interval coverage}
We wish to construct simultaneous $100(1-\alpha)\%$ one-sided upper credible intervals for $\zeta^{(l)}$, $l=1,...,L$. Denote the upper limit for the $l$th embedded DTR by $U^{(l)}$. Then, $U^{(l)}$ satisfies $$\Pr\left(\bigcap_{l=1}^L\{\zeta^{(l)}\leq U^{(l)}\}\right)=1-\alpha.$$

This is equivalent to 
    
\begin{align*}
M^{-1}\sum_{m=1}^M \mathcal{I}\left(\bigcap_{l=1}^L\{\zeta^{(l)}_{m}\leq U^{(l)}\}\right)&=
M^{-1}\sum_{m=1}^M \mathcal{I}\left(\bigcap_{l=1}^L\{\zeta^{(l)}_{(r(m,l),l)}\leq U^{(l)}\}\right)\\
M^{-1}\sum_{m=1}^M \mathcal{I}\left(\bigcap_{l=1}^L\{\zeta^{(l)}_{(r(m,l),l)}\leq U^{(l)}\}\right)&\to 1-\alpha \text{ as } M\to \infty.
\end{align*}
Let $U^{(l)}=\zeta^{(l)}_{(r_{1-\alpha}l)}$ where $r_{1-\alpha}$ is the $1-\alpha$ quantile of $r(1),...,r(M)$ where $r(m)=\max_l r(m,l)$. Then,

\begin{align*}
    M^{-1}\sum_{m=1}^M \mathcal{I}\left(\bigcap_{l=1}^L\{\zeta^{(l)}_{(r(m,l),l)}\leq U^{(l)}\}\right)&=M^{-1}\sum_{m=1}^M \mathcal{I}\left(\bigcap_{l=1}^L\{\zeta^{(l)}_{(r(m,l),l)}\leq \zeta^{(l)}_{(r_{1-\alpha}l)}\}\right)\\
    &=M^{-1}\sum_{m=1}^M \mathcal{I}\left(\bigcap_{l=1}^L\{r(m,l)\leq r_{1-\alpha}\}\right)\\
    &=M^{-1}\sum_{m=1}^M \mathcal{I}\left(\max_l r(m,l)\leq r_{1-\alpha}\}\right)\\
    &=M^{-1}\sum_{m=1}^M \mathcal{I}\left(r(m)\leq r_{1-\alpha}\}\right)\to 1-\alpha.
\end{align*}

QED.
\end{document}